\documentclass[twocolumn,showpacs,aps,prd,floatfix,nofootinbib]{revtex4}
\usepackage{graphicx}
\usepackage{epsfig}
\usepackage{times}
\usepackage{lscape}
\usepackage{rotating}
\usepackage{rotate}
\usepackage{amsmath}
\usepackage{amssymb}
\usepackage{subfigure}

\def\vdef #1{\expandafter\def\csname #1\endcsname}
\def\vuse #1{\csname #1\endcsname}

\input symbols

\long\def\inst#1{\par\nobreak\kern 4pt\nobreak
    {\it #1}\par\vskip 10pt plus 3pt minus 3pt}

\begin{document}

{%\pagestyle{empty}

% \begin{flushleft}
% CMS Analysis Document \#1\\
% Version 0 \\
% \today
% \end{flushleft}

\par\vskip 2.5cm

% Title of the paper

\title[Short Title] {SM and MSSM Higgs Boson Production: Spectra at large transverse Momentum}

\author{Urs Langenegger$^1$}
\author{Michael Spira$^2$}
\author{Andrey Starodumov$^1$}
\author{Peter Tr\"ub$^{1,2}$}

%\affiliation{$^1$ Institute for Particle Physics, ETH Z\"urich, 8093 Z\"urich, Switzerland}
%\affiliation{$^2$ Paul Scherrer Institut, 5232 Villigen PSI, Switzerland}

\affiliation{$^1$ Institute for Particle Physics, ETH Zurich, 8093
Zurich, Switzerland \\$^2$ Paul Scherrer Institute, 5232 Villigen
PSI, Switzerland}

% ----------
%  Abstract
% ----------

\begin{abstract}
\noindent 
Strategies for Higgs boson searches require the knowledge of the total
production cross section and the transverse momentum spectrum. The large
transverse momentum spectrum of the Higgs boson produced in gluon fusion
can be quite different in the Standard Model and the Minimal Supersymmetric
Standard Model. In this paper we present a comparison of the Higgs
transverse momentum spectrum obtained using the {\sc Pythia} event
generator and the {\sc Higlu} program as well as the program {\sc Hqt},
which includes NLO corrections and a soft gluon resummation for the
region of small transverse momenta. While the shapes of the spectra are
similar for the Standard Model, significant differences are observed in
the spectra of Minimal Supersymmetric Standard Model benchmark
scenarios with large $\tan\beta$.  

\end{abstract}

                                  
% For full list see http://publish.aps.org/PACS/ 
% 14.80.Bn Standard-model Higgs bosons
% 14.80.Cp Non-standard-model Higgs bosons
% 12.60.Jv Supersymmetric models

\date{\today} 

\maketitle

% ======================================================================
% The body of the paper starts here
%

% \begin{center}
% \begin{picture}(570,211) (135,-134)
% \SetWidth{0.5}
% \SetColor{Black}
% \Gluon(600,-29)(525,31){7.5}{5.43}
% \Gluon(600,-29)(525,-89){7.5}{5.43}
% \Vertex(600,-29){4.24}
% \DashLine(600,-29)(705,-29){10}
% \DashLine(285,-29)(390,-29){10}
% \Text(315,-14)[lb]{\Large{\Black{$h$}}}
% \ArrowLine(285,-29)(210,16)
% \ArrowLine(210,-74)(285,-29)
% \Gluon(210,16)(150,61){7.5}{3.93}
% \Gluon(210,-74)(150,-119){7.5}{3.93}
% \Text(630,-14)[lb]{\Large{\Black{$h$}}}
% \Text(135,-134)[lb]{\Large{\Black{$g$}}}
% \Text(135,61)[lb]{\Large{\Black{$g$}}}
% \Text(225,-29)[lb]{\Large{\Black{$t , b$}}}
% \ArrowLine(210,16)(210,-74)
% \Text(510,31)[lb]{\Large{\Black{$g$}}}
% \Text(510,-89)[lb]{\Large{\Black{$g$}}}
% \Text(435,-14)[lb]{\Large{\Black{$m_{t} \gg m_{h}$}}}
% \end{picture}
% \end{center}

\section{Introduction}

The search for Higgs bosons belongs to the most important endeavors at
the Large Hadron Collider (LHC) in order to establish experimentally
the Higgs mechanism for electroweak symmetry breaking. In the Standard
Model (SM) one isospin Higgs doublet is introduced, which leads to the
existence of one physical Higgs particle after electroweak symmetry
breaking, while the other three degrees of freedom are absorbed by the
$W$ and $Z$ bosons~\cite{Higgs:1964ia}.  In the SM Higgs sector the
only unknown parameter is the Higgs mass.  Based on triviality and
unitarity arguments its value is required to range below $\sim
800\gev$~\cite{Cabibbo:1979ay,Hasenfratz:1987eh}.  The Higgs couplings
to fermions and electroweak gauge bosons grow with the corresponding
masses ($V=W,Z$)
\begin{equation}
  g_{\f\f\h}^{SM} = (\sqrt{2} G_{F})^{1/2} m_{\f}, \qquad
  g_{VVh}^{SM} = (\sqrt{2} G_{F})^{1/2} m_{V}^2,
\end{equation}
where $G_{F} = 1.16637 \times 10^{-5} \gev^{-2}$ is the Fermi
constant. Therefore, the Higgs couplings to the
$W$ and $Z$ bosons as well as to third-generation fermions are
phenomenologically relevant, while the couplings to the first two
generations are less important.  The direct search at the LEP2
experiments excluded Higgs masses below
$114.4\gev$~\cite{Barate:2003sz}.  If the SM is embedded in a Grand
Unified Theory (GUT), the quadratically divergent radiative
corrections to the Higgs self-energy tend to push the Higgs mass
towards the GUT scale $M_{\mathrm{GUT}}\sim 10^{16}\gev$. In order to
establish a Higgs mass of the order of the electroweak scale an
unnatural fine-tuning of the counter terms is required. This hierarchy
problem remains unsolved within the SM.

The most attractive solution to the hierarchy problem is the
introduction of supersymmetry (SUSY), a novel symmetry between fermionic
and bosonic degrees of freedom~\cite{Wess:1974tw}. Due to the additional
contributions of the SUSY partners of each SM particle the quadratic
divergences in the Higgs self-energy are canceled. The hierarchy
problem is solved~\cite{Witten:1981kv}, if the SUSY particle masses are
maximally of the order of a few TeV. In the minimal supersymmetric
extension of the SM (MSSM) two isospin Higgs doublets have to be
introduced in order to preserve SUSY~\cite{Dimopoulos:1981zb} and to
render the model free of anomalies. After electroweak symmetry breaking
five Higgs bosons are left as physical particles: two ${\cal CP}$-even
neutral (scalar) particles $h,H$, one ${\cal CP}$-odd neutral
(pseudoscalar) particle $A$ and two charged bosons $H^\pm$.

At leading order the Higgs sector is determined by two independent
input parameters, which are usually chosen as the pseudoscalar Higgs
mass $M_A$ and $\tan\beta=v_2/v_1$, the ratio of the two vacuum
expectation values. The light scalar Higgs boson $h$ has to be lighter
than the $Z$ boson at leading order. This upper bound, however, is
significantly enhanced to a value of $\sim140 \gev$ due to radiative
corrections, which are dominated by top- and stop-loop
contributions~\cite{Okada:1990vk,Espinosa:1991fc}. Moreover, all Higgs
couplings are affected by the same type of corrections. The couplings
of the Higgs bosons to fermions and gauge bosons are modified by
coefficients, which depend on the angles $\alpha$ and $\beta$, where
$\alpha$ denotes the mixing angle of the two ${\cal CP}$-even Higgs
fields. The couplings, normalized to the SM Higgs coupling, are listed
in Table \ref{tb:hcoup}. An important property of these couplings is
that for large values of $\tan\beta\,$ the down(up)-type Yukawa
couplings are strongly enhanced (suppressed). The direct Higgs
searches at the LEP2 experiments have excluded neutral Higgs masses
$M_{h,H}\lsim 91.9\gev$ and $M_A\lsim 91.0\gev$ as well as charged
Higgs masses $M_{H^\pm}\lsim 78.9 \gev$~\cite{:2001xx}.
%\begin{table}[hbtp]
\begin{table}[t]
\renewcommand{\arraystretch}{1.5}
\begin{center}
\caption[]{\label{tb:hcoup} \it Higgs couplings in the MSSM to fermions
and gauge bosons ($V=W,Z$) relative to the SM couplings.}
\begin{tabular}{|lc||ccc|} \hline
\multicolumn{2}{|c||}{$\phi$} & $g^\phi_u$ & $g^\phi_d$ &  $g^\phi_V$ \\
\hline \hline
SM~ & $h$ & 1 & 1 & 1 \\ \hline
MSSM~ & $h$ & $\cos\alpha/\sin\beta$ & $-\sin\alpha/\cos\beta$ &
$\sin(\beta-\alpha)$ \\ & $H$ & $\sin\alpha/\sin\beta$ &
$\cos\alpha/\cos\beta$ & $\cos(\beta-\alpha)$ \\
& $A$ & $ 1/\tan\beta$ & $\tan\beta$ & 0 \\ \hline
\end{tabular}
\renewcommand{\arraystretch}{1.2}
\end{center}
\end{table}

At the LHC the dominant neutral SM (MSSM) Higgs production mechanisms
(for small and moderate values of $\tan\beta$ in the MSSM) are the gluon
fusion processes~\cite{Georgi:1977gs}
\begin{displaymath}
gg\to h~(h,H,A)
\end{displaymath}
which are mediated by top and bottom loops (see Fig.~\ref{fig:ggH})
and for the neutral ${\cal CP}$-even MSSM Higgs bosons $h,H$ in
addition by stop and sbottom loops with the latter contributing
significantly if the squark masses are below $\sim
400\gev$~\cite{Dawson:1996xz}. At large values of $\tan\beta$ Higgs
radiation off bottom quarks becomes competitive within the
MSSM~\cite{Spira:1997dg}.  The NLO QCD corrections to the top and
bottom loops enhance the cross sections by 50--140\% for the SM Higgs
boson~\cite{Spira:1995rr,Djouadi:1991tk} and the MSSM Higgs particles
for small values of $\tan\beta$~\cite{Spira:1995rr,Spira:1993bb},
while for large values of $\tan\beta$, where the bottom loops provide
the dominant contributions due to the strongly enhanced bottom Yukawa
couplings (see Fig.~\ref{fig:kfac}), the corrections amount to
10--60\%. It should be emphasized that the NLO QCD corrections are of
more moderate size, if the bottom loops become dominant, while they
are large in the regions of top-loop dominance.  The NNLO corrections
are known in the heavy top mass limit, which is a valid approximation
only for regions, where the top loops are dominant, i.e.,  for the SM
and the MSSM for small values of $\tan\beta$.  In this heavy top mass
limit the top loop reduces to an effective $\gluon\gluon\h$ coupling
(see Fig.~\ref{fig:ggHEff}).%
%~\cite{fn:1}.
\footnote{Note that due to the top Yukawa
  coupling the top contribution does not decouple for large top
  masses, but approaches a mass-independent value.}
This limit is expected to provide a good approximation to the exact
total cross section if $M_{\h} \lesssim 2m_t$. In the SM the maximal
deviation from the fully massive NLO result is less than $\sim 5\%$ in
this mass range, while in the MSSM it increases to $\sim 30\%$ for
$\tan\beta\sim 5$~\cite{Spira:1997dg,Kramer:1996iq}, if the full mass
dependence of the LO cross section is taken into account, while the
$K$-factor is derived in the heavy top mass limit. Compared to the
LO result, the cross section is enhanced by a factor of 1.7 to 2.3 at
NNLO~\cite{Harlander:2002wh}. Moreover, the SUSY-QCD corrections have
been calculated within the MSSM in the limit of heavy SUSY particle
and top masses.  They turn out to be large for the squark
loops~\cite{Dawson:1996xz}, while the genuine SUSY-QCD corrections,
mediated by virtual gluino and stop exchange, are ${\cal O}(5\%)$ and
thus small~\cite{Harlander:2003bb}. In the gluon fusion processes the
Higgs bosons are produced with vanishing transverse momenta at leading
order.

\begin{figure}[t]
  \begin{centering}
    \includegraphics[width=.3\textwidth]{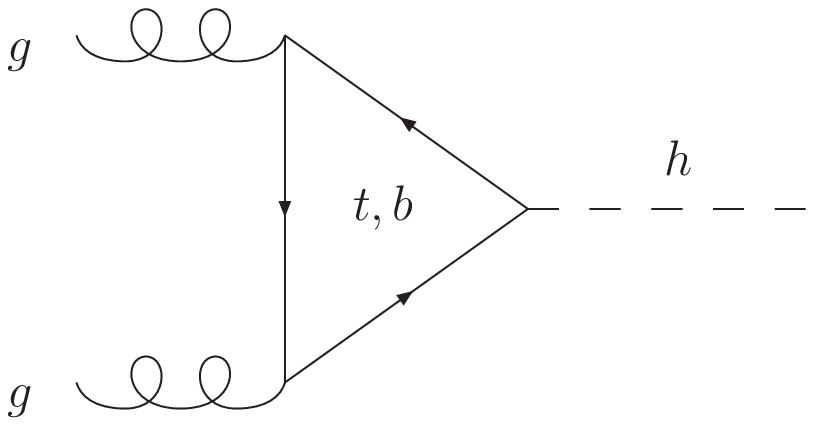}
    \caption{Leading order contribution to the SM process $\gluon\gluon \to \h$.}
    \label{fig:ggH}
  \end{centering}
\end{figure}

\begin{figure}[t]
  \begin{centering}
    \includegraphics[width=0.5\textwidth,angle=0]{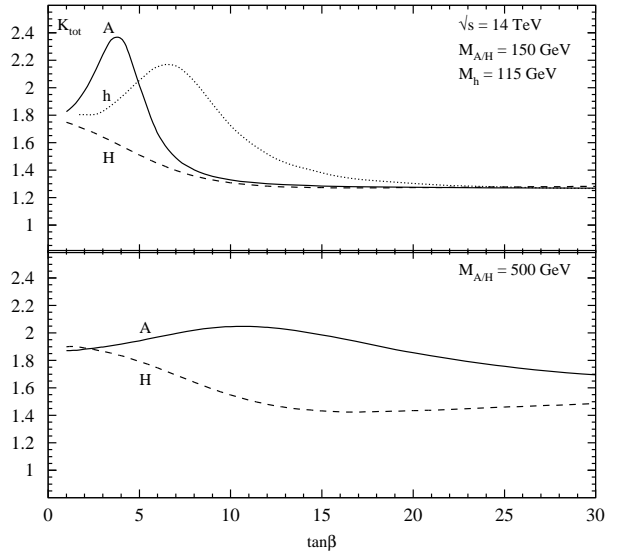}
\vspace*{-1.2cm}
    \caption{Dependence of the $K$ factors for the gluon-fusion cross
      sections on the value of $\tan\beta$. The corresponding $K$
      factors obtained by omitting the bottom loops are: $K_h=1.71$,
      $K_H=1.76~(M_H=150\gev)$, $K_A=1.78~(M_A=150\gev)$,
      $K_H=1.91~(M_H=500\gev)$ and $K_A=1.87~(M_A=500\gev)$
      independent of $\tan\beta$. CTEQ6L1 (CTEQ6M) parton
      densities~\cite{Pumplin:2002vw} are used for the LO (NLO) cross
      sections with the corresponding Higgs mass as the
      renormalization and factorization scale.}
    \label{fig:kfac}
  \end{centering}
\end{figure}

\begin{figure}[t]
 \setlength{\unitlength}{1mm}
  \begin{centering}
    \includegraphics[width=0.45\textwidth]{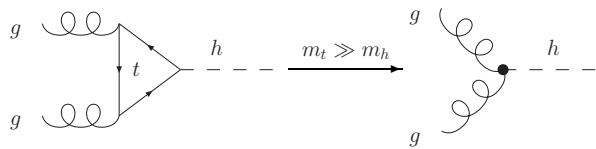}
   \put(-42.0,11.1){\vector(1,0){15}}

    \caption{Effective $\gluon\gluon\h$ coupling in the heavy top mass limit.}
    \label{fig:ggHEff}
  \end{centering}
\end{figure}

Higgs boson production at finite transverse momenta requires the
additional radiation of a gluon or quark in the gluon fusion process.
The leading order contributions to the differential gluon fusion cross
section $d\sigma / dp_T$ stem from diagrams as those in
Fig.~\ref{fig:ggHg}.%~\cite{fn:2}.
\footnote{In this letter we do not consider Higgs
  radiation off bottom quarks $gg,q\bar q\to b\bar bH$, which is of
  comparable size as the gluon fusion process for large values of
  $\tan\beta$ within the MSSM.}
\begin{figure*}[t]
  \begin{center}
  \subfigure{\includegraphics[scale=.5]{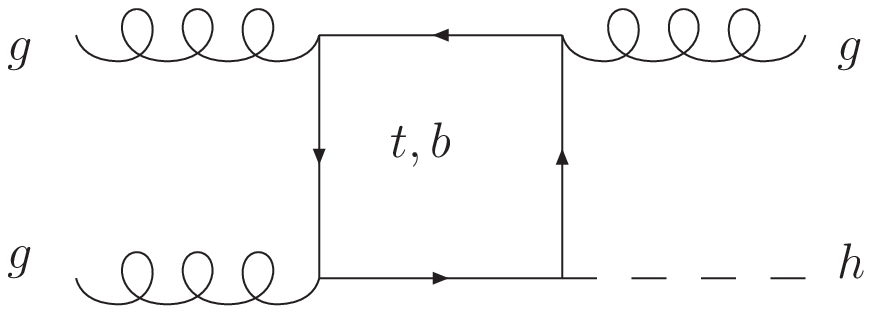}}
  \hspace{.7cm}
%   \hfill
  \subfigure{\includegraphics[scale=.5]{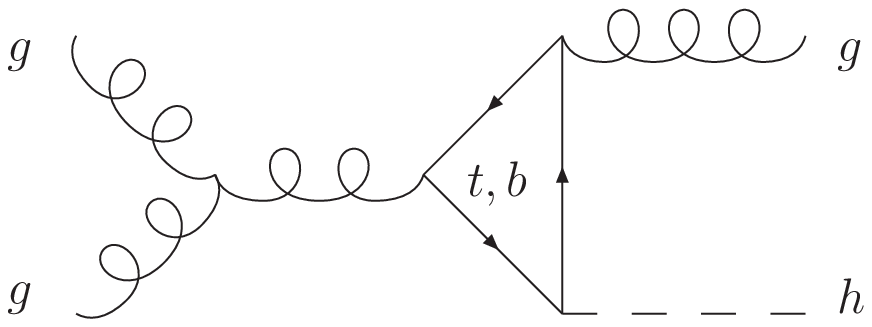}}
  \hspace{.7cm}
  \subfigure{\includegraphics[scale=.5]{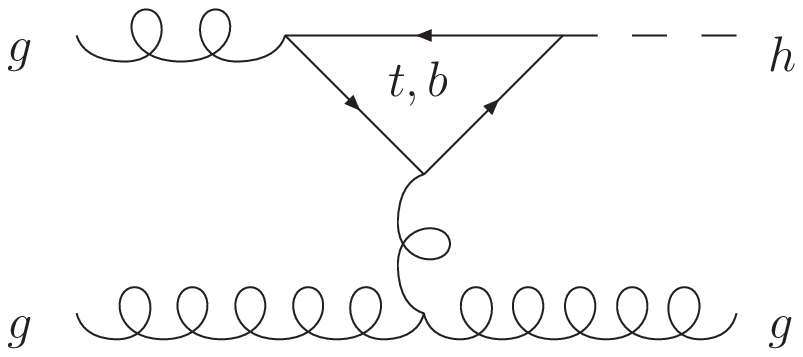}}
  \subfigure{\includegraphics[scale=.5]{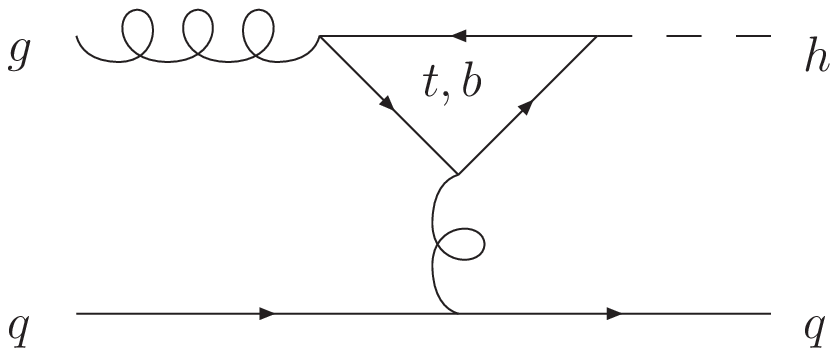}}
  \hspace{.7cm}
  \subfigure{\includegraphics[scale=.5]{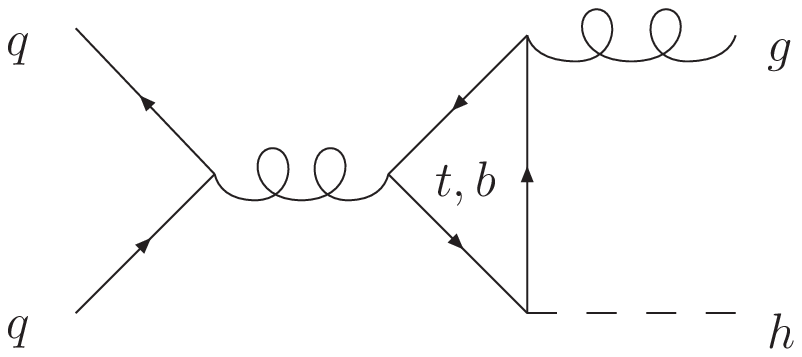}}
  \end{center}
  \caption{LO Feynman diagrams contributing to the transverse momentum
spectrum of the SM Higgs boson, mediated by Higgs couplings to gluons.}
  \label{fig:ggHg}
\end{figure*}
Analytical results can be found in~\cite{Ellis:1987xu}. NLO
contributions to these expressions are again only known in the heavy
top mass limit~\cite{Schmidt:1997wr}, which serves as an approximation
for $M_h,p_T\lesssim m_t$. Since the bottom mass is small, this
approximation is not expected to work in MSSM regions, where the
bottom loop contributions are significant. At low transverse momentum
($p_T \ll M_h$) multiple soft gluon emission spoils the validity of
fixed order calculations. To obtain reliable results in this region,
contributions of all orders have to be taken into account by resumming
the large logarithmic terms $\ln^n(m_{\h}^2/p_T^2)$. These
computations have been performed at leading logarithmic (LL),
next-to-leading logarithmic (NLL)~\cite{Catani:1988vd} and
next-to-next-to-leading logarithmic (NNLL) level~\cite{higgsptnnll}.
The NNLL calculation was performed in the heavy top mass limit. To
obtain reliable predictions for the differential cross section over
the whole $p_T$ range, the resummed results are matched to fixed order
calculations at large transverse momentum. This is for example done in
the {\sc Hqt}~\cite{Bozzi:2005wk} program (for more details on the
programs see below), results of which are shown in
Fig.~\ref{fig:fixedMatched}.  The heavy top mass limit is only valid
as long as $p_T \lesssim m_t$.  For larger values this approximation
tends to overestimate the cross section~\cite{Ellis:1987xu}. This can
be seen in Fig.~\ref{fig:higluPythiaSM}, where the SM spectrum was
computed with {\sc Higlu}~\cite{Spira:1995mt} for the exact top mass
and an infinite top mass.

\begin{figure}[hbtp]
  \begin{centering}
    \includegraphics[width=0.5\textwidth]{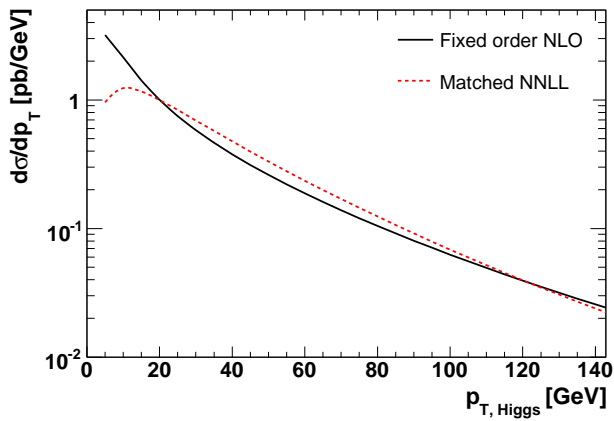}
    \caption{Comparison of the NLO differential cross section with the
NNLL result, matched to the NLO result for SM Higgs bosons with mass
$M_h=115\gev$ in the heavy top mass limit. The renormalization and
factorization scales are chosen as the transverse mass.}
    \label{fig:fixedMatched}
  \end{centering}
\end{figure}

The development of strategies for Higgs boson searches at the LHC
requires reliable estimates of the total and the differential cross
section. The latter can potentially be used to discriminate signal
from background, especially at large transverse momenta, where many
background processes are suppressed~\cite{Mellado:2004tj}.  In this
letter we compare the transverse momentum distributions of the SM and
MSSM Higgs bosons obtained with the programs {\sc
  Pythia}~\cite{Sjostrand:2001yu}, {\sc Higlu}~\cite{Spira:1995mt} and
{\sc Hqt}~\cite{Bozzi:2005wk}. In {\sc Pythia} the large transverse
momentum spectra have the same shape in the SM and the MSSM. A
crosscheck of the results of {\sc Pythia} at large transverse momenta
is performed with the program {\sc Higlu}~\cite{Spira:1995mt}. In the
SM the spectrum obtained by {\sc Higlu} is softer than that of {\sc
  Pythia}, because {\sc Pythia} uses the heavy top mass approximation.
The assumption in {\sc Pythia} of similar transverse momentum shapes
in the SM and the MSSM is not correct in general, especially in
regions of bottom-loop dominance~\cite{Field:2003yy}.  In some MSSM
benchmark scenarios large discrepancies between the two programs are
found.

\section{Numerical results and discussion}

If not mentioned otherwise, all results are given for a Higgs boson mass
of $115 \gev$ and a value of $\tan\beta = 30$. For the numerical studies the following
programs have been used:

\begin{itemize}
  \item[-] {\sc Pythia} v6.227~\cite{Sjostrand:2001yu} \\
 The {\sc Pythia} process $\gluon\gluon \to \h$ uses the matrix
 element at order $\alpha_s^2$, i.e., the Higgs boson is produced at
 rest.  The Higgs boson receives its transverse momentum only by
 initial state radiation, which is added to the hard interaction. The
 initial state radiation is added in such a way that the large $p_T$
 spectrum is matched to that of the process $\gluon\gluon \to
 \h\gluon$. All calculations implemented in {\sc Pythia} are performed
 in the heavy top mass limit. We have used the default scale choices.

  \item[-] {\sc Higlu}~\cite{Spira:1995mt} \\
 {\sc Higlu} is a program to compute the total Higgs production cross
 section via gluon fusion at NLO in the SM as well as in the MSSM. It
 can also be used to calculate the differential cross section at the
 same order in $\alpha_s$ (i.e. at LO), although this option is not
 documented. To the best of our knowledge, {\sc Higlu} is the only
 freely available program which allows to compute the differential
 spectrum in the MSSM with the full heavy quark mass dependence. The
 renormalization and factorization scales are chosen as the transverse
 mass $M_T = \sqrt{M_h^2+p_T^2}$.

  \item[-] {\sc Hqt}~\cite{Bozzi:2005wk} \\
 {\sc Hqt} can be used to compute the $p_T$ spectrum of the Higgs
 boson produced in gluon fusion at LO and NLO. The spectra are
 calculated in the heavy top mass limit, but they can be normalized to
 the exact total cross sections. {\sc Hqt} only performs calculations
 in the SM. It includes a soft gluon resummation, thus providing a
 reliable and finite prediction in the limit of small transverse
 momentum in contrast to the purely perturbative result implemented in
 {\sc Higlu}, which diverges for $p_T\to 0$. The renormalization and
 factorization scales are chosen as the transverse mass $M_T =
 \sqrt{M_h^2+p_T^2}$.

\end{itemize}

If not mentioned otherwise, CTEQ5L parton densities were used in 
{\sc Pythia}, while in {\sc Higlu} and {\sc Hqt} CTEQ6M parton
densities were used~\cite{Pumplin:2002vw}.

\subsection{Standard Model}

As shown in Fig.~\ref{fig:higluPythiaSM} good agreement between the
spectra of {\sc Pythia} and {\sc Higlu} is obtained in the SM, if {\sc
  Higlu} is used in the heavy top mass limit. The shapes of the two
curves coincide reasonably well at large $p_T$. At low $p_T$ the {\sc
  Higlu} result diverges, because {\sc Higlu} does not perform any
resummation. 
While good agreement is achieved in the heavy top mass
limit, significant deviations for large values of $p_T$ are observed
when taking into account finite quark mass effects in {\sc Higlu}:
the cross section at large $p_T$ is overestimated by the heavy top
mass approximation~\cite{Ellis:1987xu} as can be inferred also from
Fig.~\ref{fig:higluPythiaSM}. Thus any analysis requiring Higgs
bosons at large transverse momenta has to be adjusted accordingly.

\begin{figure}[t]
  \begin{centering}
    \includegraphics[width=0.5\textwidth]{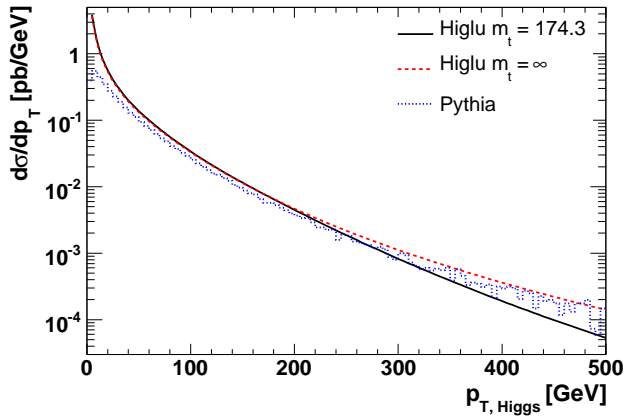}
    \caption{Comparison of the differential spectrum in the SM generated
by {\sc Higlu} and {\sc Pythia} for $M_h=115\gev$. The renormalization 
and factorization scales in {\sc Higlu} are chosen as the transverse
mass.}
    \label{fig:higluPythiaSM}
  \end{centering}
\end{figure}

\subsection{Minimal Supersymmetric Standard Model}

In the MSSM the comparison between {\sc Pythia} and {\sc Higlu} was
carried out for the four different benchmark scenarios proposed
in~\cite{Carena:2002qg}, i.e., the $m_h^{\mathrm{max}}$, the
\textit{no-mixing}, the \textit{gluophobic Higgs} and the
\textit{small} $\alpha_{\mathrm{eff}}$ scenario.

In {\sc Pythia} the shape of the spectra is the same for the SM and all
MSSM scenarios, only the normalization changes. This is true even though
by switching on supersymmetry in {\sc Pythia}, the couplings of the
quarks to the Higgs boson change compared to the SM.  As in the SM, the
Higgs particle is produced at rest and then initial state showers are
added to the hard scattering. This shower is using a splitting kernel
that is convoluted with the LO $\gluon\gluon\to\h$ matrix element,
thereby ensuring that the shower reproduces the
$\gluon\gluon\to\gluon\h$ matrix element at large $p_T$ values by a
proper matching procedure. This convolution of the splitting kernel with
the hard interaction is performed in the heavy top mass limit in the SM
and MSSM. Therefore the spectrum shapes coincide in both cases. This can
be clearly seen in Fig.~\ref{fig:pythiaShapes}, where all spectra
produced by {\sc Pythia}, scaled to the corresponding total SM cross
sections, are shown.

\begin{figure}[t]
  \begin{centering}
    \includegraphics[width=0.5\textwidth]{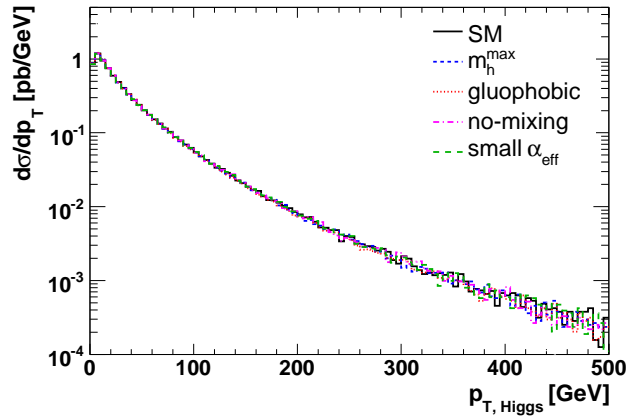}
    \caption{{\sc Pythia} spectra for the SM and the four benchmark
scenarios scaled to the SM total cross section. The light scalar Higgs
spectra are shown for $M_h=115\gev$.}
    \label{fig:pythiaShapes}
  \end{centering}
\end{figure}

\begin{figure*}[t]
  \begin{center}
  \subfigure{\includegraphics[scale=.43]{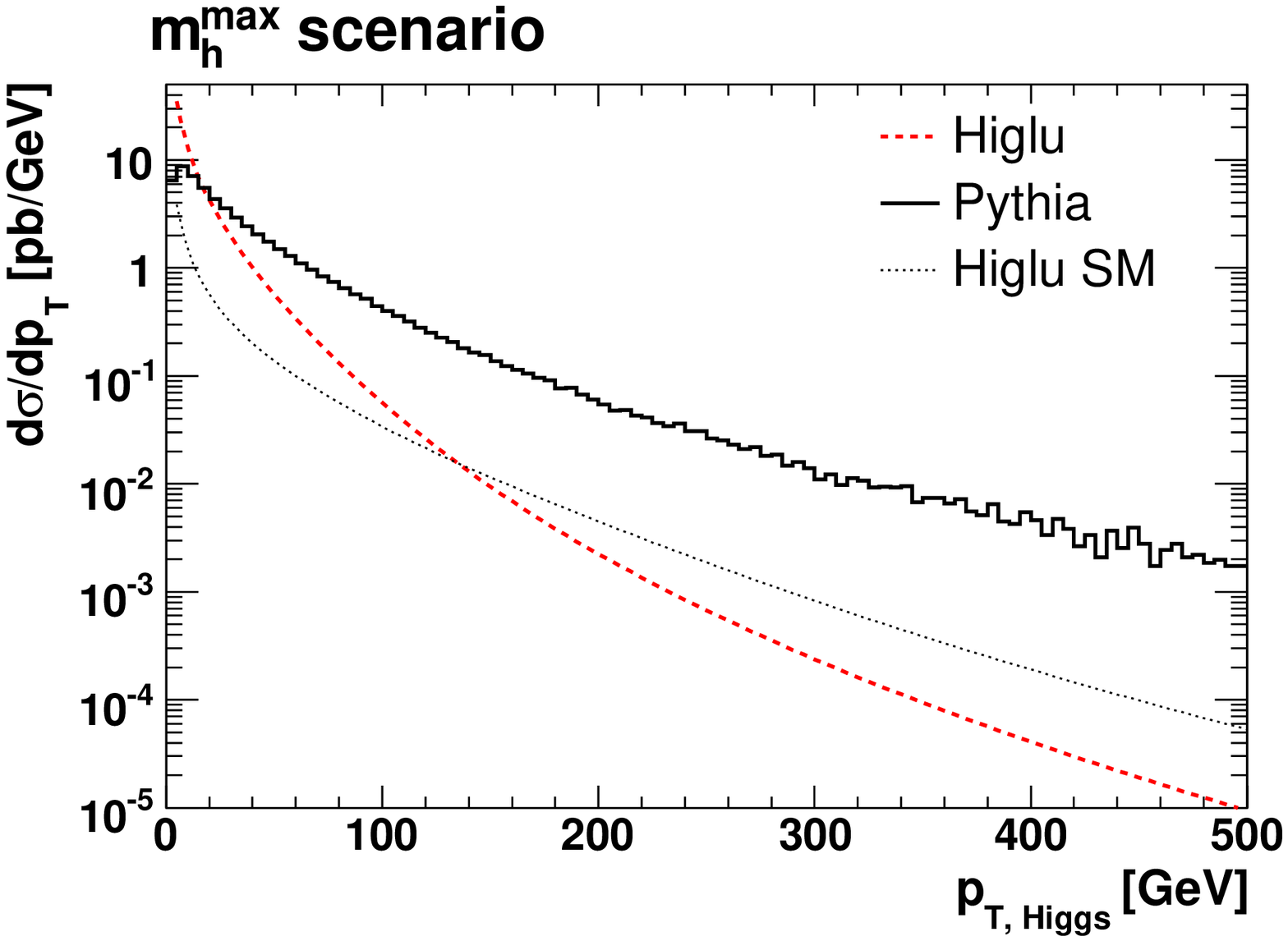}}
  \hspace{.5cm}
  \subfigure{\includegraphics[scale=.43]{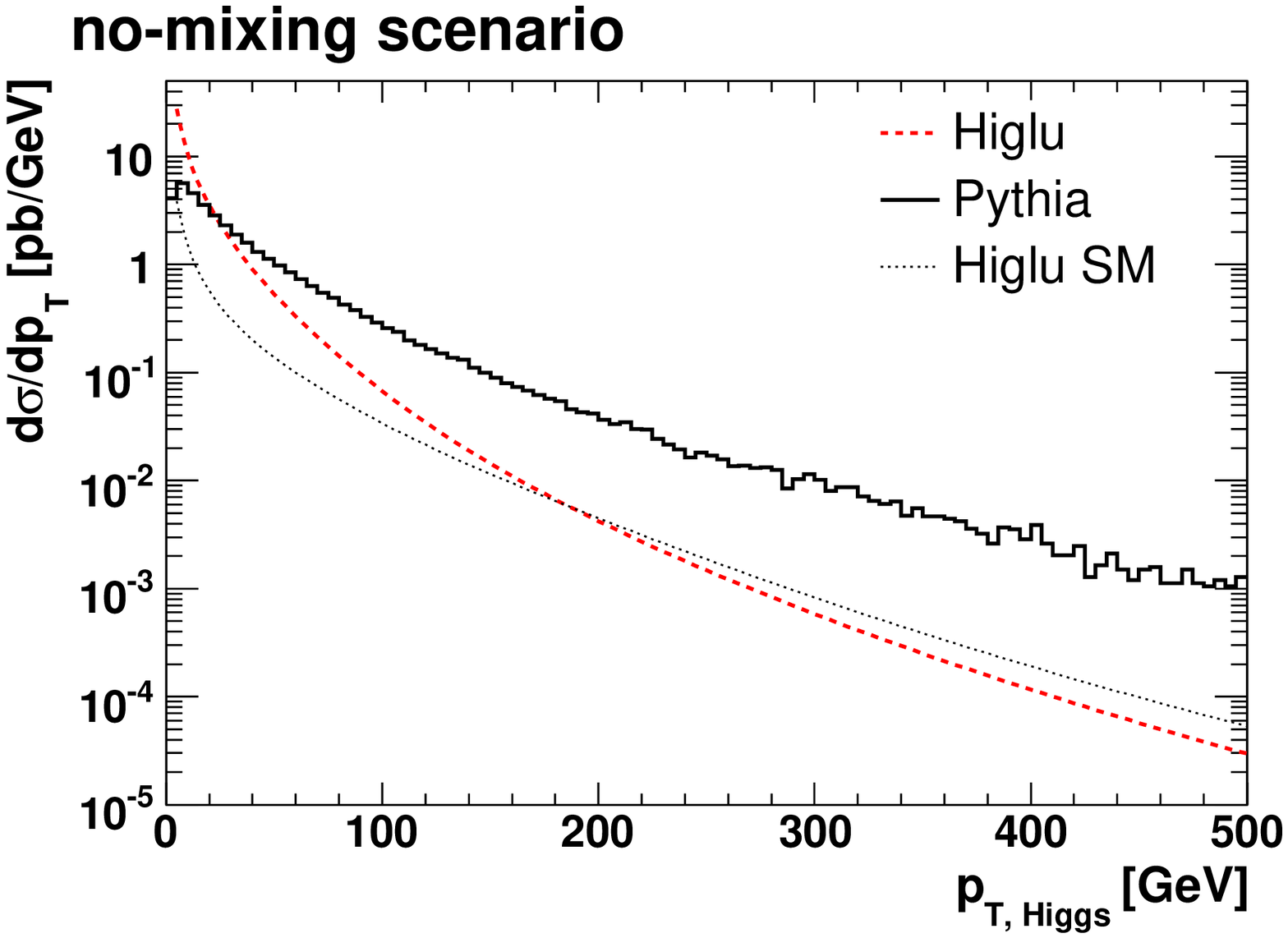}}
  \subfigure{\includegraphics[scale=.43]{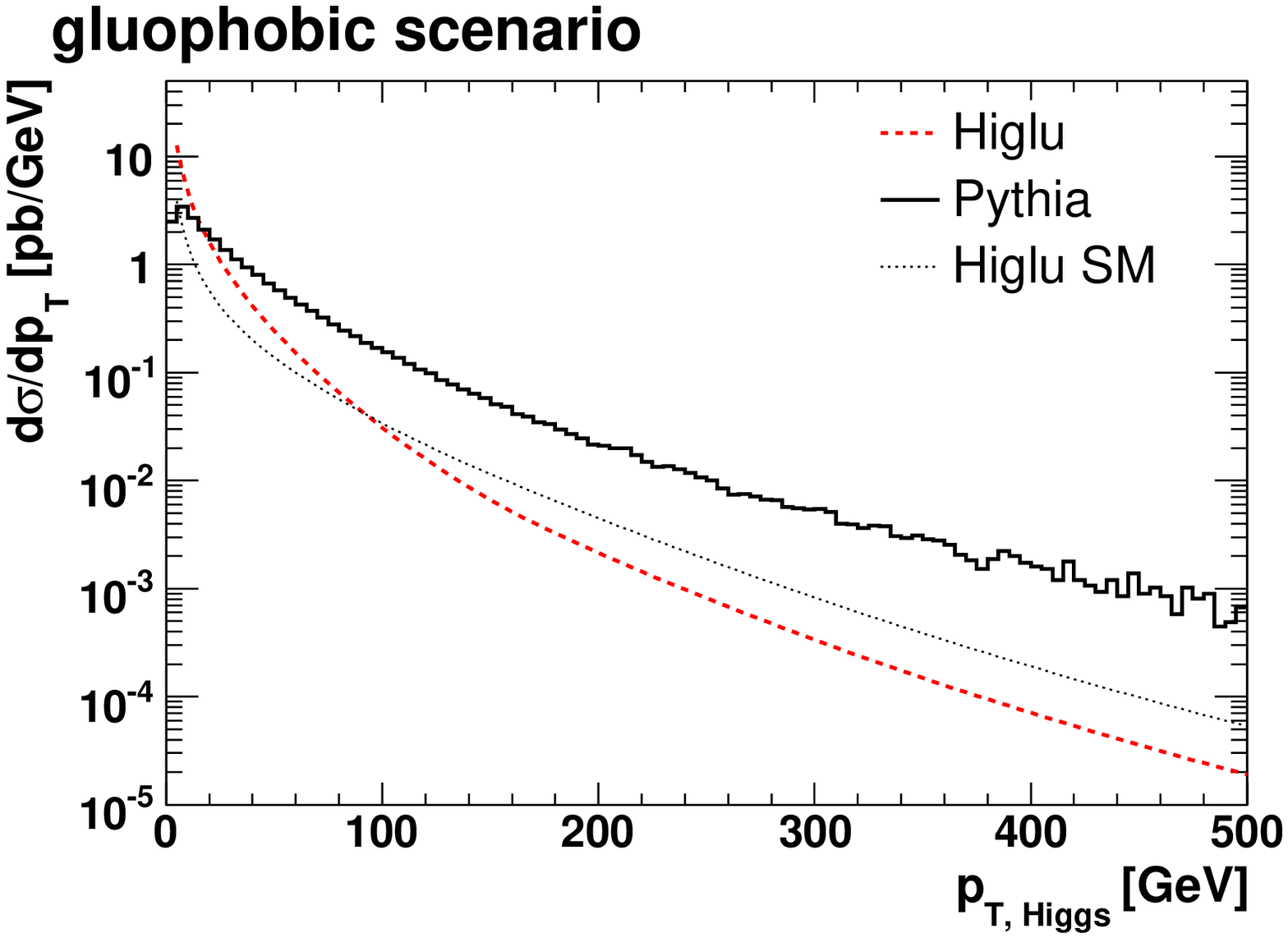}}
  \hspace{.5cm}
  \subfigure{\includegraphics[scale=.43]{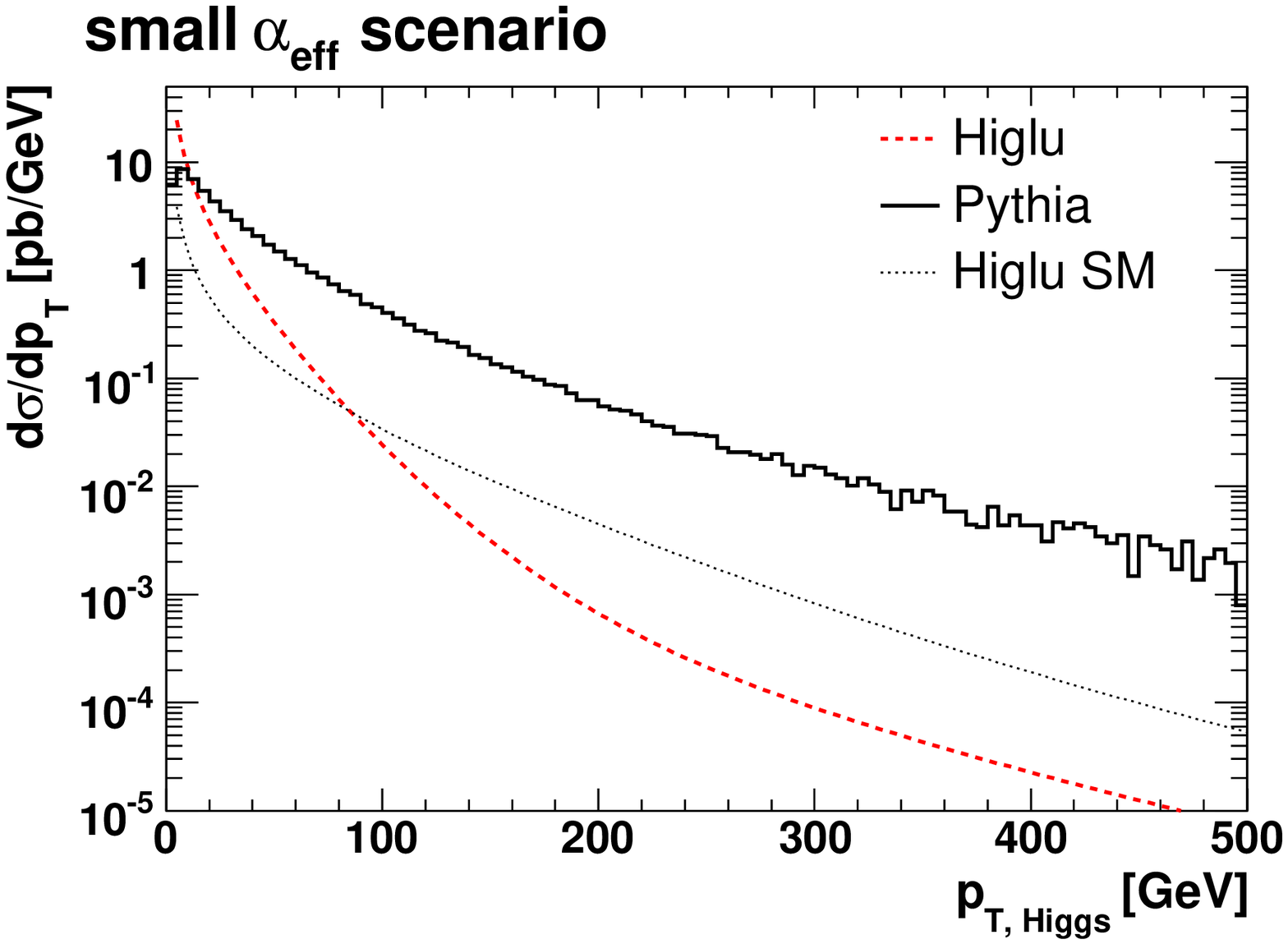}}
  \end{center}

    \caption{Comparison of the differential spectra generated by {\sc
Higlu} and {\sc Pythia} in the four benchmark scenarios for the light
scalar Higgs boson with $M_h=115$ GeV.}
    \label{fig:higluPythia}
\end{figure*}

In contrast to this, {\sc Higlu} shows differently hard spectra for the
benchmark scenarios, as can be inferred from Fig.~\ref{fig:higluPythia}.
All spectra turn out to be softer than the SM spectrum. Taking the {\sc
Pythia} spectrum as the base of a physics analysis, the Higgs signal can
be overestimated by more than an order of magnitude for $p_T \gtrsim 100
\gev$. The softness of the spectra can be traced back to the fact, that
the bottom quarks yield the main contribution to the differential cross
section for large $\tan\beta$. Table~\ref{tab:couplings} lists the
couplings for the four benchmark scenarios. In the $m_h^{max}$ scenario,
the bottom loop contribution is enhanced by a factor of $1.65 \times
10^4$ compared to the top-loop contribution alone. A lighter quark in
the loop generates a softer spectrum as can be read off
Fig.~\ref{fig:mtopDependency}. The same effect is also visible in the
spectra of the other scenarios. It is particularly large in the \textit{small} 
$\alpha_{\mathrm{eff}}$ scenario.

\begin{table}[!hbt]
  \begin{center}

    \caption{LO cross sections and Higgs Yukawa couplings to up- and
down-type quarks in the four benchmark scenarios computed with {\sc
Higlu} for $\tan\beta = 30$ using CTEQ6L1 parton densities for the light
scalar MSSM Higgs boson with $M_h=115\gev$.}

    \vspace{0.1in}
    \begin{tabular}{|l|c|c|c|c|}
      \hline
                                &$\sigma_{LO}$           [pb]&$g^\h_\d$
&$g^\h_\u$           &$(g^{\h}_\d/g^{h}_\u)^2$ \\
      \hline
        SM                      &21.8                   &1.00
&1.00                   &1.00\\
      \hline
        $m_h^{\mathrm{max}}$             &200                    &26.0
&0.202                  &$1.65 \times 10^4$\\
      \hline
        {\it no-mixing}         &158                    &24.6
&0.544                  &$2.05 \times 10^3$\\
      \hline
        {\it gluophobic}        &71.7                   &16.6
&0.800                  &$4.32 \times 10^2$\\
      \hline
        {\it small} $\alpha_{\mathrm{eff}}$  &141                    &18.3
&-0.464                 &$1.56 \times 10^3$\\
      \hline
    \end{tabular}
    \label{tab:couplings}
  \end{center}
\end{table}

\begin{figure}[t]
  \begin{centering}
    \includegraphics[width=0.5\textwidth]{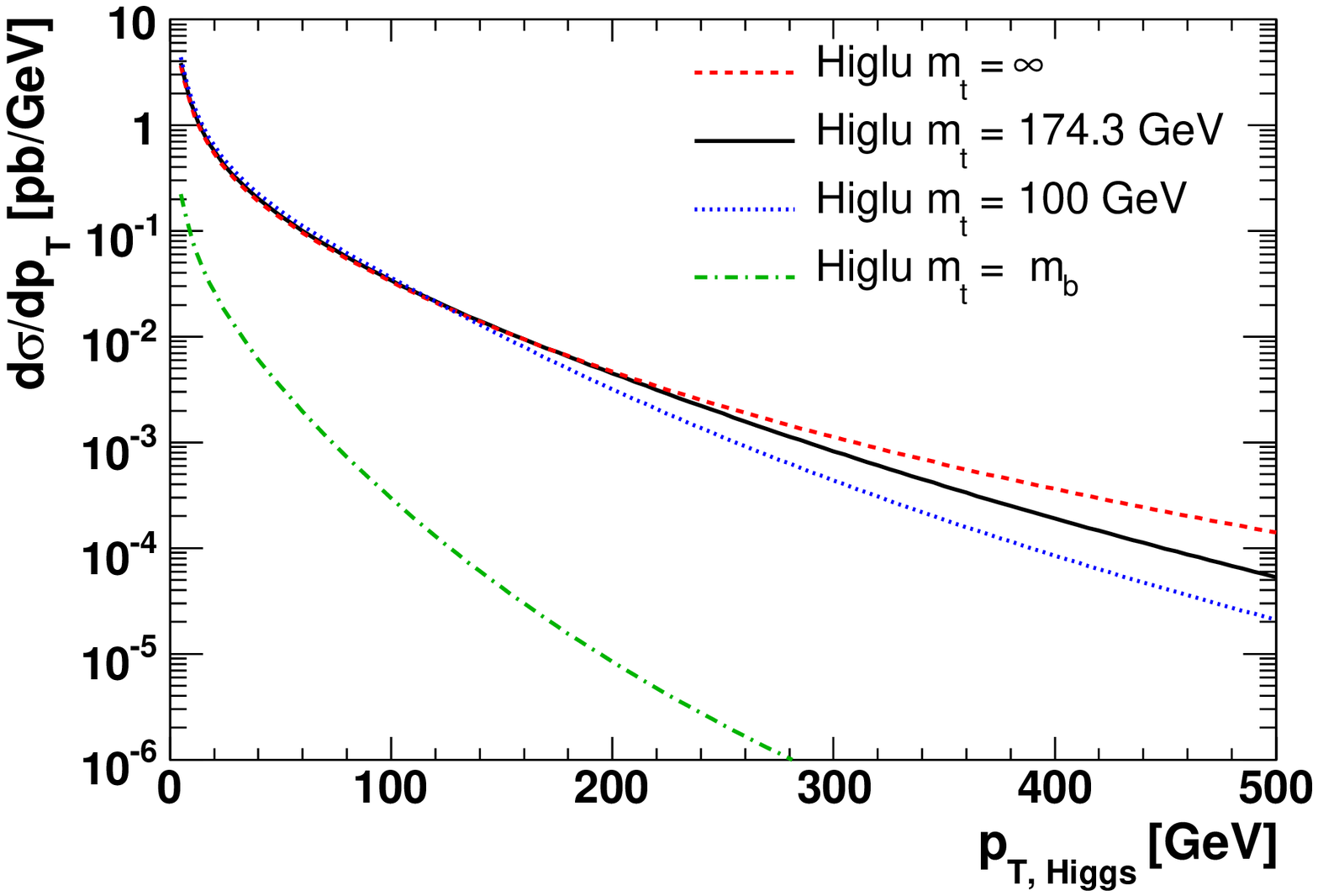}
    \caption{Differential cross section in the SM for different top
masses and $M_h=115\gev$.}
    \label{fig:mtopDependency}
  \end{centering}
\end{figure}

The effect of two other parameters, the Higgs mass and $\tan\beta$, on
the large $p_T$ spectrum is shown in Figs.~\ref{fig:mhiggsDependency}
and \ref{fig:tanbDependency}. If the mass of the light Higgs boson
approaches its upper limit (which corresponds to $M_A \rightarrow
\infty$), the spectrum becomes SM-like. The same happens if $\tan\beta$
is lowered as presented in Fig.~\ref{fig:tanbDependency}, since in both
limits, $M_A\to \infty$ and small $\tan\beta$, the Higgs Yukawa
couplings become similar to the SM couplings.
Fig.~\ref{fig:scaleDependency} shows the variation of the spectra if the
renormalization and the factorization scales are varied by a factor two
around the transverse mass $M_T=\sqrt{M_h^2+p_T^2}$. The transverse
momentum distributions changes by $\sim \pm 40\%$. This variation can be
taken as a lower bound of the theoretical uncertainty at LO in analogy
to the total inclusive gluon fusion cross section.

\begin{figure}[t]
  \begin{centering}
    \includegraphics[width=0.5\textwidth]{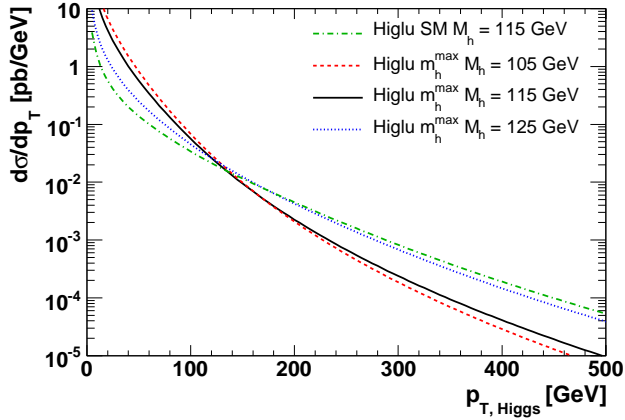}
    \caption{Differential light scalar MSSM Higgs cross sections for
different Higgs masses in the $m_h^{max}$ scenario.}
    \label{fig:mhiggsDependency}
  \end{centering}
\end{figure}

\begin{figure}[t]
  \begin{centering}
    \includegraphics[width=0.5\textwidth]{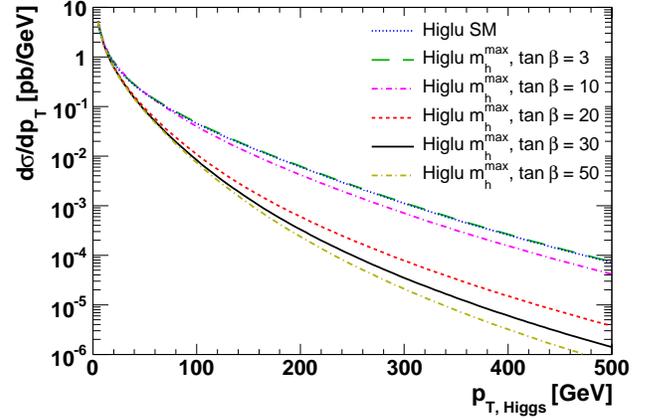}
    \caption{Differential light scalar Higgs cross section for different
values of $\tan\beta$ and $M_h=115\gev$. All spectra are rescaled by
the ratio of the total LO SM cross section and the corresponding total
LO cross section of the scenario.}
    \label{fig:tanbDependency}
  \end{centering}
\end{figure}

\begin{figure}[t]
  \begin{centering}
    \includegraphics[width=0.5\textwidth]{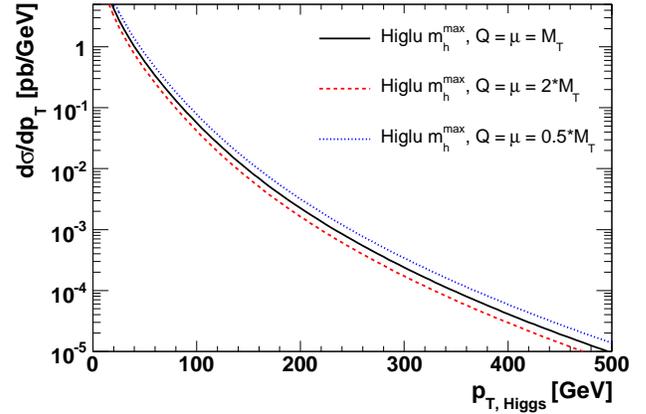}
    \caption{Light scalar transverse momentum spectra in the $m_h^{max}$
scenario for different renormalization ($\mu$) and factorization ($Q$)
scales and $M_h=115\gev$.}
    \label{fig:scaleDependency}
  \end{centering}
\end{figure}

\subsection{Best estimate for the large $p_T$ spectrum}

The aim of this subsection is to get the best possible prediction for
the differential cross section at large transverse momentum. It is clear
from the preceding discussion that the following conditions have to be
met:

\begin{itemize}
  \item The full quark mass dependence has to be taken into account
  \item For large $\tan \beta$ the bottom quark loops must not be omitted
  \item The calculation has to be performed at the highest possible order in $\alpha_s$.
\end{itemize}

By using {\sc Higlu}, the two first conditions are automatically met.
The third one is not completely fulfilled, because {\sc Higlu}
performs the calculation only at third order in the strong coupling
constant. Presently there are no calculations at order $\alpha_s^4$
fulfilling the first condition, since at this order the differential
cross section is only known in the heavy top mass limit.  In the
meantime, the following improvement to the {\sc Higlu} result is the
only possibility.  From the comparison of the differential cross
section in the heavy top mass limit at LO and NLO a $p_T$ dependent
$K$-factor can be extracted, which then can be applied to the exact LO
result. This procedure has been carried out for the example of the
$m_h^{max}$ scenario. The $p_T$ dependent $K$-factor was computed with
{\sc Hqt} and then applied to the LO {\sc Higlu} spectrum. The result
is shown in Fig.~\ref{fig:bestEstimate}. The figure shows the {\sc
  Hqt} spectra at LO and NLO, from which the $K$-factor was extracted,
the {\sc Higlu} result at LO as well as the scaled {\sc Higlu}
spectrum. For comparison the {\sc Pythia} result is also given.
However, it should be noted that this approximation will be valid only
for small values of $\tan\beta$, where the top loops are dominant, and
not too large transverse momenta, while its validity for large values
of $\tan\beta$, where the bottom loops become dominant, is not clear,
before a fully massive NLO calculation is available for large
transverse momenta.

\

\begin{figure}[t]
  \begin{centering}
    \includegraphics[width=0.5\textwidth]{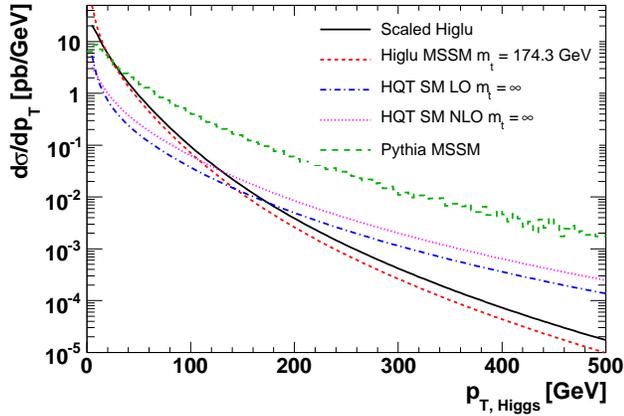}
    \caption{Best estimate for the large transverse momentum spectrum of
the light scalar Higgs boson in the $m_h^{max}$ scenario for $M_h=115\gev$.}
    \label{fig:bestEstimate}
  \end{centering}
\end{figure}

\section{Conclusion}
The large transverse momentum spectrum of the Higgs boson produced in
gluon fusion was investigated with different programs. While the
predictions of {\sc Pythia} and {\sc Higlu} agree reasonably well
within the SM, significant differences were found in the MSSM at large
values of $\tan\beta$. Compared to {\sc Higlu}, {\sc Pythia}
overestimates the differential cross section in some scenarios by more 
than one order of magnitude for $p_T > 100 \gev$. The reason for this is, that {\sc
Pythia} is always working in the heavy top mass limit, where the quark
loop is replaced by an effective coupling of the two gluons to the
Higgs boson. In this way, the shape of the {\sc Pythia} spectrum is
not sensitive to changes of the Higgs Yukawa couplings, which modify
the relative weight of the top and bottom quark loop contributions.

\section{Acknowledgments}
U.L. and P.T. acknowledge discussions with Stefano Frixione and Zoltan
Kunszt.  This  research was partially supported by  the Swiss National
Science Foundation (SNF).

% -- The form for references is directly from SPIRES. 

% %%%%%%%%%%%%%%%%%%%%%%%%%%%%%%%%%%%%%%%%%%%%%%%%%%%%%%%%%%%%%%%%%%%%%%

\end{document}